# Knotted proteins: Tie Etiquette in Structural Biology

Ana Nunes and Patrícia FN Faísca*
Departamento de Física and BioISI - Biosystems and Integrative Sciences Institute, Faculdade de Ciências, Universidade de Lisboa, Campo Grande, Ed. C8, Lisboa, Portugal

*Corresponding author:
PFNF, email: pffaisca@fc.ul.pt

**Abstract**

A small fraction of all protein structures characterized so far are entangled. The challenge of understanding the properties of these knotted proteins, and the why and the how of their natural folding process, has been taken up in the past decade with different approaches, such as structural characterization, *in vitro* experiments, and simulations of protein models with varying levels of complexity. The simplest among these are the lattice Gō models, which belong to the class of structure-based models, i.e., models that are biased to the native structure by explicitly including structural data. In this review we highlight the contributions to the field made in the scope of lattice Gō models, putting them into perspective in the context of the main experimental and theoretical results and of other, more realistic, computational approaches.

*Introduction*

Proteins are chains of amino acids that fold into specific three-dimensional shapes, known as native conformations, to become biologically functional. Finding the native structure of all encoded amino acid sequences is an on-going task, but more than a hundred thousand protein native conformations are already freely available in the Protein Data Bank (PDB) (1), in the form of Cartesian coordinates for each atom in the protein. It has been known for some time that some of these native structures have their backbone knotted (2,3).

A topological knot (4) is a closed curve in space, and all its continuous deformations are by definition equivalent knots. Any such curve may be represented by a planar projection with a finite number of double crossings, together with the specification of the over- and under-strand at each crossing. A reduced knot diagram is one of these planar representations for which the minimum number of crossings is attained. This number, called the crossing number, is therefore a topological invariant. The simplest knot is the circle, with crossing number 0, and the simplest non-trivial knots are the trefoil ($3_1$ in the Alexander–Briggs notation that will be used throughout) and the figure-eight ($4_1$) knots, with crossing numbers three and four, respectively. There are two different knots with crossing number five, the cinquefoil ($5_1$) and the twisted-three ($5_2$) knot, and three different knots with crossing number six. The number of knots increases very rapidly with crossing number from then on, but there are no known examples of protein folds with crossing number larger than six.

In rigor proteins do not form topological knots because they are open chains. Therefore, it is more accurate to say that protein conformations embed physical (or open) knots, although in many cases the termini can with little ambiguity be connected with an external arc to form a closed loop **(Fig. 1A-B)**. Specific computational methods have been developed to determine if a given protein conformation is knotted, and to make it amenable to knot invariant calculations (in most cases the Alexander polynomial) so as to find its knot type. An important example is the Koniaris–Muthukumar-Taylor (KMT) algorithm developed in 2000 by Taylor (5), which extends to a wide range of protein conformations the application of the Koniaris–Muthukumar method (6) developed for ring polymers **(Fig. 1C-D)**. Alternative methods to determine the knotting state of proteins based on different loop closure procedures have since then been proposed (7-11). More recently, the concept of knotoids introduced by Turaev (12) set the basis for a new method to study the entanglement of open protein chains (13), further refined in (14), and now available as an open access software tool (15).

Independently of these methodological advances, the overall picture of the topological properties of the known native conformations is by now well established (16,17). The systematic application of knot detection methodologies based on the KMT algorithm to all available protein entries in the PDB revealed that about 1% correspond to knotted proteins (10,16). Although the trefoil is by large extent the most common knot type, it is possible to find a few proteins with more complex knots, including the Stevedore knot $6_1$ (with six crossings on a planar projection) (18). An interesting variation amongst tangled proteins is that of the slipknot, in which one (or both (19)) protein termini adopts a hairpin-like conformation that threads a loop formed by the remainder of the chain (20). There are ~500 slipknotted proteins in the PDB (16) .

These numbers mean that, in a sense to be precised in the following sections, knotted proteins are uncommon, but the reasons for why the protein universe avoids knots have not yet been fully established. It is possible that local geometric features of proteins that do not favor tangling of the backbone were selected for, but it is also likely that knots were partly removed by evolution due to their unfavorable effect on protein folding. To clarify whether knots were selected against as a global property or indirectly as a result of selection operating on protein structure at the local level it is necessary to understand how knotted proteins fold.

If it is already challenging to determine how ‘regular’ proteins fold, it is even more formidable to do it in the case of knotted proteins, which have distinctively complicated native structures. But a complete solution to the protein folding problem must necessarily encompass the solution to this new exciting folding puzzle (21). Therefore, perhaps not surprisingly, during the last decade knotted proteins became a favorite theme within the protein folding community. Several research groups worldwide, both experimental and theoretical have been making significant efforts to solve this tangling puzzle using simulated environments, both in vitro and *in silico* (reviewed in (22)), and first steps have been taken in order to determine how this intricate process occurs inside the living cell (23,24).

Although the overwhelming majority of protein structures in the PDB are unknotted, some proteins are knotted - in contrast with the RNA structures in the PDB, for which the incidence of knots is practically negligible (25). Whatever the selection mechanism that operated, it spared a few structures. The realization that knotted proteins exist motivated the search for the functional advantages that knots may convey to their carriers. The analysis of specific model systems has put forward the idea that knots could enhance thermal (20,26) or mechanical stability (27), just to mention a few examples. However, a universal role for knots in proteins has not yet been identified.

Here we review a set of results, which address the questions outlined above in the scope of molecular simulations. This will not be a comprehensive review article on knotted proteins. We focus on structured-based models (including our own work on lattice models of protein folding) and integrate these views in a broader perspective. There are recent, thorough review articles on the subject that complement the present work (28-32), and the interested reader is invited to consult them for a comprehensive perspective on these interesting matters.

*Structure-based models*

In the 1970s Nobuhiro Gō proposed the now famous Gō potential for protein folding (33). According to the Gō potential, protein folding energetics is driven only by the intramolecular interactions present in the native structure. Therefore, only those interactions can contribute to energetically stabilize a folding chain. The Gō potential was originally developed in the context of a lattice representation **(Fig. 2A-C)**. A lattice representation is one in which the amino acids are reduced to beads of uniform size and placed on the vertices of a two- or three-dimensional regular lattice, with the lattice spacing representing the peptide bond. Lattice models, such as the one used in the original Gō model, retain the fundamental features of polypeptide chains (connectivity, excluded volume etc.) and can be used to explore fundamental aspects of the folding process (e.g. how folding rate depends on the chain length etc.). The combination of the Gō potential with off-lattice representations (including C-alpha and full atomistic resolutions) allows exploring the folding mechanism of specific model systems **(Fig. 2D-F)**.

The Gō model is the archetypal structure-based model (SBM). A SBM is one that imposes a native bias by explicitly including structural data (34) (Fig. 2). In off-lattice representations the atomic coordinates derived from NMR, X-ray or Cryo-EM must be used to construct the intramolecular potential. SBMs became particularly popular in the beginning of this century after the experimental realization that native ‘topology’ (geometric properties of the native configuration measured by the contact order (35) or other metrics (36,37)) is a major drive of the two-state folding transition of small, monomeric proteins (see also (38)). On the theoretical side, they are consistent with the energy landscape framework, which envisions proteins as minimally frustrated heteropolymers with a smooth funnel-like energy landscape biased toward the native state (39-41). SBMs can be combined with Monte Carlo methods, and Molecular Dynamics protocols. They offer a clear advantage over more realistic force fields (i.e. those traditionally used in classical Molecular Dynamics such as the GROMOS or AMBER force fields (42)) because of their lower computational cost, especially when combined with replica-exchange simulations (43) and other sampling protocols (44) designed to accelerate conformational search and relaxation towards equilibrium. Despite their simplicity, SBMs have been successful in the study of protein folding (34,45-47) and, more recently, in exploring other phenomena involving proteins such as aggregation (48-50), protein fold switches (51), and phase separation (52). Clearly, SBMs, and in particular Gō models, will not be able to correctly model misfolding processes leading to compact non-native states and, more generally, regions of the free energy landscape where non-native interactions play an important role (53-55). Interestingly, a seminal study by Shakhnovich and co-workers, which was the first to explore the folding mechanism of knotted trefoil (56), reported that a simple C-alpha representation combined with a Gō potential would not be able to efficiently fold the protein: the process had necessarily to be assisted by non-native interactions. This claim sparked an interesting discussion on the role of non-native interactions in the folding of knotted proteins and triggered a series of studies that contributed to clarify their folding mechanism and stimulate experimental investigations as we outline below.

*Knotting physics and why are knotted proteins rare*

The starting point to understand the role, if any, of topological complexity in biopolymers must be the comparison with the frequency and type of knots we expect to find in homopolymers of comparable length. Before embarking on an overview of the state-of-art on the folding mechanism of knotted proteins we propose to discuss some aspects underlying the physics of knotting in a more general perspective, and then narrowing down our discussion towards the specific case of proteins. This will contribute to clarify the question of rarity of knotted proteins and motivate the need to understand their folding mechanism. Interestingly, many of the results we will overview have been obtained in the scope of simple models including lattice representations.

The self-avoiding random walk (SAW) on a three-dimensional lattice is perhaps the simplest model for a real polymer, one in which excluded volume and stiffness are minimally represented, favoring local entanglement. Indeed, it was conjectured by Frisch, Wassermann and Delbruck that sufficiently long polymers must be knotted (57), and rigorously proved by Sumners and Whittington (58) that almost all sufficiently long self-avoiding walks on the three-dimensional simple cubic lattice contain a knot.

Application to real polymers required some information about the typical length of knotted SAWs, as well as the extension to models including more physical features. This prompted a large body of numerical studies based on self-avoiding polygons, both on-lattice (59,60) and off-lattice with variable thickness (6,61,62). These results, reviewed and expanded in (63), show that probability of the trivial knot in a ring polymer with $N$ segments $P(N)$, is well approximated by an exponential function, $P(N) \sim \exp(-N / N_0)$, so that there exists indeed a model dependent knotting length scale $N_0$. On lattice, $N_0$ is of the order of $10^5$. Off-lattice models yield values of $N_0$ that are very sensitive to excluded volume, ranging from ~300 for random polygons (no excluded volume) to ~10000 when the polymer thickness is about 1% of the edge length.

A study of maximally compact conformations obtained from Hamiltonian walks on cubic lattices found an exponentially decaying probability $P(N)$ with knotting length scale $N_0$ ~196 (64), showing that chain compactification favors knot formation even more than ignoring excluded volume. The fact that proteins are compact polymers is therefore a key ingredient to gauge their topological properties through numerical studies of abstract models, and since most proteins in the PDB have a few hundred amino acids, the value found for $N_0$ in this study already suggested that the low abundance of knotted proteins cannot be ascribed to chance. Moreover, knots in these abstract models should in general be harder to obtain than the knots usually found in proteins, the latter being typically shallow, i.e., they become unknotted by removing a small number of amino acids starting from one of the termini.

More recently, statistical analysis of large samples of random polygons bounded within spheres of variable radius was employed (65-67) to explore in this context the effect of confinement on knotting probability and knot type, as well as on the relation between knot type and average geometrical properties, such as total curvature and total torsion.

Progress towards more realistic physical models was obtained with off-lattice numerical simulations of a simplified model of polyethylene (68), represented by chains of interacting monomers connected by springs, parametrized so as to reproduce experimental data. At high temperatures, the entropic term of the free energy dominates, favoring the swollen or coil state. By contrast, at low temperatures, the energetic contribution of the interaction potential dominates, favoring compact conformations – the globule state. Simulations of the system in these two regimes with chain lengths up to $N = 1000$ recapitulate the two complementary scenarios found the abstract models mentioned above. Knots are indeed rare in the coil phase, with only around one percent of configurations knotted for $N = 1000$, and common in the globule state, with about half the conformations knotted for $N \sim 600$.

Another way of endowing a linear or a ring polymer with physical features (besides excluded volume) is to introduce its rigidity through a tunable bending energy. A systematic survey of the interplay between chain length, bending rigidity and knotting (69) confirmed the low abundance of knotted conformations for swollen polymers (less than 2% for $N = 500$) and found a surprising effect. As chain flexibility increases, the equilibrium conformations become more compact, but their knotting probability is strongly non-monotonic, exhibiting a pronounced maximum for intermediate bending rigidities. This counterintuitive phenomenon is shown to be due, in a more subtle way than in the previous example, to the competition between the energetic and the entropic contributions.

The studies mentioned above also include estimates for the probability $P_K(N)$ of specific non-trivial knots of type K as a function of chain length and other model parameters, with the trefoil, knot $3_1$ being by far always the most abundant, in agreement with PDB data.

Perhaps not surprisingly given the specificity of each protein's amino acid sequence and spatial arrangement, the application these ideas and results to biopolymers has not been as illuminating and conclusive in the case of proteins and RNA as in the case of DNA (29).

Because it is easier to produce and analyze knots in DNA than in other polymers, most experiments on knotting involve DNA (29), and a wealth of experimental data has been collected for the past thirty years. It was shown as early as in 1993 that the description of a random self-avoiding semiflexible chain of a given thickness reproduced both qualitatively and quantitatively the features of free DNA (70). In this work, the effective diameter of the DNA double helix was determined by comparing experimental results with computer simulations of knotting probability, and the results for the diameter in NaCl solutions at different concentrations were found to be in agreement with independent theoretical predictions based on polyelectrolyte theory.

DNA molecules in the cell are in general in a crowded or confined physical environment that differs drastically from that of free molecules in good solvent conditions. Compactification should increase the knotting probability, and systematic simulations of flexible bead-spring rings confined in a sphere of radius R (71) quantified this effect by finding a scaling function of R and N for the ratio of the probabilities of the trivial knot in the free and in the confined system. Further studies extended the model to include chain thickness and bending rigidity, with realistic parameter choices for DNA, and the equilibrium knot spectrum – the probabilities of formation of the different knot types – was estimated (72). The goal of these theoretical efforts was to reproduce the main features found in a series of experimental data available for bacteriophage DNA packed in the capsids (reviewed in (29) and (73)), but one intriguing aspect of the observed knot spectrum remained unexplained. The bacteriophage genome is very tightly packed in the capsid, displaying an extremely high knotting probability (~95%) together with a dominance of very complex knots (most of the knots have crossing number larger than 10). Another characteristic of DNA arrangement in the capsid is that the frequency among simple knots of certain knot types, such as $4_1$ or $5_2$, is much smaller than that of other knot types, such as $3_1$ and $5_1$. The knot $4_1$ is achiral, i.e., equivalent to its own mirror image, while the other three knot types lack this symmetry. The abundant knot types $3_1$ and $5_1$ are torus knots, i.e., knots that can be drawn on the surface of a torus. This apparent bias of the knot spectrum towards torus knots and against achiral knots could not be reproduced with the inclusion of chain thickness and bending rigidity alone. A successful description was obtained later in (73), by introducing a DNA-DNA interaction potential that translates into a preferred “twist angle” between DNA segments that are close in space.

In contrast with DNA, the incidence of knots in RNA has only recently been systematically explored (25), with a very clear and surprising conclusion: the knotting frequency of RNA in the PDB is virtually zero. Physical and biological causes for the absence of knotted RNA are still being debated (74).

The first systematic comparative study of the geometric and topological properties of natural proteins and compact random homopolymers, as a function of their length, is due to Lua and Grosberg (9). The topological analysis showed that knots in proteins are orders of magnitude less frequent than in random polymers of comparable length, compactness, and bending rigidity. The geometrical analysis showed that, although the overall geometry of the conformations is statistically close to random, there are significant differences at the local (<10 residues) and intermediate (between 15 and 40 residues) levels. On a local scale, due to the formation of secondary structure (α-helices and β-chains), proteins are more stretched than random. On an intermediate scale, protein chains tend to fold back on themselves and crumple, a feature that reflects the most common arrangements of consecutive secondary structure elements and was known to be consistent with the suppression of knots. Thus, a connection was established between knot avoidance, a global property, and the local properties of protein chains.

This explanation of why there are so few knotted proteins leads to the question of why are there knotted proteins at all, and what is special about them. Pioneering experimental work in knotted proteins, Jackson and co-workers found that proteins YibK and YbeA unfold reversibly upon denaturation, and the native knot type appears spontaneously and reproducibly in the same protein location (21,75), inspiring a systematic study of all knotted, versus unknotted, protein structures in the PDB, together with their sequences (76). The analysis revealed in several knotted structures the presence of relatively short segments whose removal results in structurally similar though unknotted configurations, suggesting that there are localized ‘knot-promoting’ regions within knotted proteins. It also showed that there is little similarity, at the sequence level, between knotted and unknotted proteins.

The relation between knottedness and sequence was explored in the scope of an on-lattice HP model (77). The HP model represents a protein as self-avoiding chain of amino acids of two types only, ‘hydrophobic’ (H) or ‘polar’(P), and the only interaction present is the attraction of neighboring H-H pairs (78). Sampling large ensembles of native conformations for different random and designed sequences showed that the mean unknotting probability, $P_0$, is about 0.46, slightly larger than for homopolymers, but there is a large variability in sequence space, with $P_0$ ranging from 0.3 to 0.6 in a set of 100 random sequences and examples of designed sequences with $P_0$ as high as 0.897 and as low as 0.114.

Taken together, these results strengthen the idea that knot abundance in proteins has to be understood in the light of evolution.

*First insights into the folding of knotted proteins via molecular simulation*

Despite remarkable advances in experimental studies on the folding of knotted proteins (23,79-85), it is not yet possible to obtain a structurally resolved picture of the knotting mechanism. Furthermore, the difficulty in creating unfolded ensembles of unknotted conformations through chemical denaturation (86) creates an additional challenge to explore this process through in vitro studies. A main advantage of molecular simulations is to obtain predictions (often with atomic detail) that can be experimentally tested, leading to a deeper understanding of these intricate processes. Sometimes, the results from simulations also contribute to prompt and design novel experiments, and, on the other hand, experiments can be used to refine simulations (87).

The first computational study addressing the problem of understanding how a knotted protein folds, was carried out by Shakhnovich and co-workers in 2007 (56). They combined Langevin Dynamics simulations with a C-alpha representation to explore the folding mechanism of protein YibK (PDB ID: 1j85), which embeds a trefoil knot. This is a bacterial alpha-beta protein of the methyltransferases (MTases) family, which, together with YbeA (PDB ID: 1vh0), have been extensively studied to shed light on the folding of knotted proteins. YibK comprises 156 residues, and the knotted core, i.e., the minimal segment that contains the knot, is located 77 residues away from the N-terminus and 39 residues away from the C-terminus (Fig. 1(AC)). YibK classifies, therefore, as a deeply knotted protein; in contrast, if the removal of a few residues is enough to untie the carrier protein, the corresponding knot classifies as shallow. Shakhnovich's study provided the very first view on the remarkably intricate knotting mechanism, including the structure of the knotting step, and timing of knot formation. Indeed, it predicted that to get knotted, part of the polypeptide chain of YibK must first arrange itself in a loop (the so-called knotting loop, which is not necessarily in the native conformation), which is subsequently threaded by the C-terminus **(Fig. 3A)**. The authors pointed out that in most of the cases threading of the C-terminus proceeds directly, but in some trajectories the C-terminal region is, instead, arranged in a hairpin-like conformation **(Fig. 3B)**. Furthermore, knotting in their model system occurs predominantly late, in near-native (and therefore compact) conformations (with fraction of native contacts $Q \sim 0.8$).

The most striking conclusion of this seminal study concerns the energetics of the knotting step. Indeed, it was reported that when only native interactions are considered (as in a pure Gō potential), the conformations are too compact to allow for threading events, and none of the 100 attempted folding runs was successful. Non-native interactions are essential for folding, and in order to observe 100% folding efficiency (i.e. all attempted folding trajectories lead to native conformations) threading must be assisted by a set of specific non-native attractive interactions established between a stretch of residues within the knotting loop (86-93) and the residues pertaining to the C-terminus.

*Insights into the knotting mechanism from molecular simulations*

A follow-up study by Sulkowska *et al.* analysed the folding process of YbiK and YbeA within a similar simulation set-up (88). In line with the results reported by Shakhnovich, only a small (~1-2%) percentage of successful folding trajectories was recorded for a pure Gō potential. It was found that in the folding runs that lead to the native state, the proteins populate a conformation that embeds a slipknot. The slipknot arises when the C-terminal region arranges itself in a hairpin like conformation and threads a partially structured region that contains the knotting loop. This intermediate conformation bears resemblance with that reported by Shakhnovich and co-workers for the less likely folding pathway of the modified Gō potential.

In the wake of these studies several contributions followed, exploring other model systems and using alternative simulation strategies. In particular, inspired by Sulkowska's results, Faísca *et al.*(89) decided to explore in the scope of Monte Carlo simulations the folding of a designed lattice protein embedding a shallow trefoil in its native structure. Like Sulkowska, they also used a pure Gō potential to model intramolecular interactions. All of the thousands attempted folding runs were successful, and the occurrence of both knotting mechanisms was detected. However, direct threading was the most frequent one. The higher efficiency of the lattice model may result from the fact that the analyzed system is considerably smaller (41 residues long), and the knot is shallow, which facilitates direct threading of the shorter knot tail. By measuring the knotting frequency (the fraction of knotted conformations) against the folding probability, $P_{fold}$ (an accurate kinetic estimator of folding progression)(90), it was found that the formation of the knot is mainly a late folding event, occurring with high probability (≥ 0.6) in conformations with $P_{fold} > 0.8$. However, a subsequent study revealed that the folding efficiency decreases dramatically upon tethering the protein to a chemically inert surface through a neutral linker (91). In particular, if surface tethering occurs at the bead that is closer to the knotted core, the folding rate becomes exceedingly (i.e. about two orders of magnitude) slow, and the protein is no longer able to find the native structure in all the attempted folding trajectories. The knotting frequency keeps a negligible value throughout the folding process, increasing abruptly to ~1 when the protein is nearly folded, with more than 80% of its native contacts formed. These results predict that the mobility of the terminus closest to the knotted core is critical for efficient folding, which, in turn, highlights the importance of a knotting mechanism that is based on a threading movement of the shorter knot tail. Interestingly, Lim and Jackson provided the first experimental evidence for a knotting mechanism based on the threading of the C-terminus for proteins YbiK and YbeA (24). Indeed, as predicted by the simulations mentioned above, hampering the threading of the C-terminus, i.e., the one closer to the knotted core, also results in a dramatic reduction of the *in vitro* folding rate for these model systems.

Škrbić and co-workers used Monte Carlo simulations of a C-alpha model to investigate the folding process of protein AOTCase (PDB ID: 2g68) (92). This is a 332 amino acid long chain that contains a deep trefoil knot located 173 residues away from the N-terminus and 81 residues away from the C-terminus. When a pure Gō potential is used to model protein energetics, the knotting frequency is always negligible throughout the folding process. However, when non-native interactions are included, the knotting propensity raises considerably along the folding process. This behavior may result from the fact that non-native interactions energetically stabilize partially folded conformations, therefore lowering the free energy barrier of the entropically costly knotting step (85). Furthermore, for this protein, the knotting step also consists of a direct threading movement of the C-terminus, in line with that observed by Shakhnovich for YibK (56).

A follow-up study looked into the folding mechanism of MJ0366 (PDB ID: 2efv), using an advanced simulation technique known as the dominant reaction pathway, combined with a realistic force field with implicit solvent (93). MJ0366 is the smallest (82 amino acids) knotted protein in the PDB embedding a shallow trefoil knot, located 10 residues away from the C-terminus. The authors reported that knotting typically occurs at a late folding stage, in conformations with $Q=0.90$, and is almost invariably based on a direct threading mechanism, in line with the observations for the shallow knotted system (89) mentioned above. These results are somehow different from those previously reported by Noel *et al*. for the same model system when analyzed within the scope of a full atomistic model combined with a Gō potential (94). Indeed, in this case, it appears that side-chain packing leads to the population of an intermediate state with about 25% of native contacts formed (i.e. $Q\sim0.25$), which is structurally characterized for having the native knotting loop formed. Two knotting pathways are observed: one, based on a slipknotting mechanism, dominates below the folding transition temperature, $T_f$, (i.e. the temperature at which native and unfolded states are equally populated under equilibrium conditions) while the other, based on a direct threading of the C-terminus becomes equally likely at $T_f$. Interestingly, recent experimental studies based on NMR hydrogen-deuterium analysis also suggested the population of an on-pathway intermediate, although considerably more structurally consolidated than the one predicted by simulations (95). 15 selected conformations containing the slipknot were subsequently used as starting conformations in Molecular Dynamics simulations in explicit solvent with the amber99sb force field (96). These conformations had 10 (out of 15) residues already threaded and had yet to thread another 5 residues to form the native knot. Five of these conformations reached the native state in 450 ns, adding to the plausibility of a knotting mechanism based on a spliknotted conformation in real proteins. Remarkably, it appears that the native contacts that thread the terminus through the knotting loop are conserved in the realistic force field, supporting the use of structure-based models to study complex folding processes.

The results outlined above indicate that in trefoil proteins the knotting mechanism is generally based on a threading movement of the C-terminus. The latter can be direct, or, instead, involve the population of a conformation with a slipknot. Evidence for the existence of a knotting mechanism based on threading was reported by Jackson for protein YibK, in recent *in vitro* experiments (24). Alternative mechanisms have been proposed in lattice models (spindle mechanism)(91), and for specific proteins (e.g. loop flipping was observed in simulations of MJ0366 (92), and proposed for the protein hypothetical RNA methyltransferase (PDB ID: 1o6d) following *in vitro* experiments (82)).

Although there is some lack of consensus regarding the predominance of direct threading over slipknotting, simulations from different groups, and using different simulation strategies and models, all agree with the timing of the knotting step, i.e., that knotting occurs towards late folding, in structurally consolidated conformations. However, folding of proteins YibK and YibA was observed in a cell-free translating system, mimicking *in vivo* conditions, when one (either the C- or the N-) terminus, or both termini simultaneously were fused to stable (ThiS) domains (24). Because ThiS domains are bulky, the knotting step of fused YibK and YibA proteins should involve a threading movement of the full ThiS domain. The latter can only occur early in the folding process, when the knotting loop is sufficiently loose to allow for the passage of large structures. Because knotting is entropically costly (85) such mechanism may seem unlikely in the beginning of the folding process since there are not yet enough native interactions established to decrease the free energy barrier to knotting. However, and at least for the YibK model system, one possibility to circumvent this challenge is that knotting occurs co-translationally (i.e. during protein synthesis in the ribosome), via a slipknottted conformation (97), as we discuss later in this article.

*Knot type and knotting mechanism*

While most studies addressing the folding of knotted proteins have been based on knotted trefoils, it is clear that a comprehensive understanding of the folding and knotting process should include the study of model systems embedding other knot types. The number of theoretical articles in the literature addressing this problem is scarce. To the best of our knowledge Virnau and co-workers were the first to explore the folding of protein DehI, embedding a deep $6_1$ knot (18), and Soler *et al*. (98) were the first to study the folding of a lattice model system with a shallow $5_2$ knot (98). More recently, Sulkowska and co-workers, provided the first off-lattice results for protein Ubiquitin C-terminal Hydrolase L1 (UCH-L1) (PDB ID: 3irt) embedding a shallow $5_2$ knot in its native structure (99).

DehI is a homodimer, formed by two 130 residues long monomers connected by a shorter linker region to complete the $6_1$ knotted topology. The folding route of Dehl was explored via Molecular Dynamics simulations with a coarse-grained C-alpha model combined with the Gō potential (18). Out of 1000 runs, six trajectories folded into the native knotted state with very similar routes, suggesting that DehI folds via a simple mechanism: two large loops are formed early in the simulations by twists of the partially unfolded protein, and the six-fold knot is created later in a single movement when either one of the loops flips over the other. Experimental analysis of the equilibrium unfolding of DehI by chemical denaturation (100) suggests that DehI first unfolds in to a monomeric intermediate state with 42% of the native secondary structure still formed, which then unfolds into another intermediate state with only 21% of the secondary structure content but a similar level of compactification, and finally total unfolding, and potentially unknotting, is achieved, accompanied by a total loss of secondary structure. Thus, in DehI knotting precedes the formation of most of the secondary structure, but this experimental analysis could not ascertain whether the actual knotting mechanism is as proposed by Virnau and co-workers.

*Soler et al.* considered a lattice protein with chain length $N = 52$, designed by hand to have a shallow $5_2$ knot in its native structure (98). The folding mechanism of the $5_2$ knot shares with that of shallow the $3_1$ knot ($N=41$) the occurrence of a threading movement of the chain terminus that lays closer to the knotted core. However, in sharp contrast to what is observed for the knot $3_1$, knotting occurs particularly late during folding for the $5_2$ knot (e.g. when $Q$~0.7 the knotting frequency is 0.82 for the lattice trefoil but is only 0.12 for the twisted-three).

The Ubiquitin C-terminal Hydrolase (UCH) is a family of knotted proteins with the $5_2$ knot, and with sub-chains that form the $3_1$ knot. Experimental studies of UCH-L3(101) show two parallel folding pathways, each associated with the formation of a rapidly populated intermediate state. Using a structure-based C-alpha model, Sulkowska and co-workers (99) did a computational study of three members of the UCH family that is consistent with the main trends of the experimental results. They also found two, topologically distinct pathways. One in which the N-terminus is structured last and the $5_2$ knot is formed directly from the unknot, and another in which the C-terminus is structured last and a $3_1$ knotted intermediate appears.

*Folding properties of knotted proteins*

Perhaps one of the most remarkable features of the overall folding behavior of knotted proteins regards the persistence of knotted conformations in chemically denatured states, a finding that was originally reported by Mallam and Jackson for proteins YibK and YibA (86). Recent experiments by Capraro and Jennings for a protein of the same family, the hypothetical RNA methyltransferase (PDB ID: 1o6d), whose native structure also embeds a deep trefoil knot, revealed that in order to untie it, it is necessary to use highly denaturing conditions during several weeks (82). Since original experimental studies (79,101,102) were not aware of this resilience to untie upon chemical denaturation, the first kinetic measurements provided folding rates in line with those exhibited by regular proteins of the same size, and were interpreted accordingly, i.e., knotted proteins fold in about the same timescale. However, they were based in folding processes that started from denatured, but knotted, conformations. As shown through molecular simulations, conformations with a small fraction of native contacts ($Q$~$0.2$-$0.3$) can fold remarkably fast (more than one order of magnitude faster) than unknotted conformations, if they keep the knotting loop (103). This behavior changes drastically when folding trajectories start from unfolded and unknotted conformations. Faísca *et al.* were the first to compare the folding rate of a knotted lattice protein with a control system (i.e. an unknotted counterpart which is obtained by minimally modifying the backbone connectivity of the knotted system that serves as a template)(89), and predicted that knotted proteins fold considerably slower, especially below the folding transition temperature, $T_f$. This is in line with in vitro results obtained by Yeates for a cleverly engineered protein, known in the literature as 2ouf, which is designed to embed a $3_1$ knot. The latter was shown to fold 20-times slower than a protein that was designed to have a similar tertiary structure but to be unknotted (84). In a subsequent study Soler *et al.* also provided compelling evidence that folding rate decreases dramatically as knot depth (103) and knot complexity increases (98). Further results from lattice (98,103) and off-lattice simulations (104,105) confirmed the same trend, i.e., that knotted proteins fold slower. An important definitive proof that knotted proteins are indeed slow folders, came from *in vitro* experiments carried out by Mallam and Jackson on YibK and YibA. They used a cell free expression system (i.e. an experimental set up containing only the elements necessary for protein translation) to synthesize YibK and YibA as a strategy to assure that the folding process started from unfolded, and unknotted, conformations. They found that YibK and YibA fold spontaneously without populating misfolded states. However, the folding process that starts from denatured conformations exhibits a rate constant that is ~3 (YibA) to ~35 (YibK)-fold greater than that exhibited by newly translated conformations, which was taken as an indication that knotting is the rate-limiting step (23). This conclusion applies if newly synthesized proteins are indeed unknotted. This may, however, not be necessarily the case for YibK since there is theoretical work predicting that knotting may occurs co-translationally (97) .

More recently, single molecule experiments accessed the effect of knot formation on the folding rate of protein ubiquitin C-terminal hydrolase isoenzyme L1, UCH-L1, (PDB ID: 3irt) (106). This protein is 223 residues long and features a shallow $5_2$ knot in its native structure. It was shown that folding from an unknotted denatured state is one order of magnitude slower than from knotted unfolded states, indicating that for this family of knotted proteins, knotting is also the rate limiting step.

A major reason why knotted proteins are slower folders is related with the phenomenon of backtracking; the term was introduced by Onuchic and co-workers to describe the process of breaking and re-establishing specific native contacts (107). It is believed that backtracking is prevalent in tangled proteins because knotting is a highly ordered process, whereby the polypeptide chain must arrange itself in a succession of specific conformations (e.g. a conformation with the knotting loop must form first that is subsequently protruded by the chain terminus). Failure in achieving the correct order of events leads to malformed knots, or other misfolded conformations, that must be unfolded to allow for productive knotting and folding. In order to probe the importance of backtracking in the folding of knotted proteins, Soler *et al.* introduced the so-called structural mutations (SM)(98). SMs disrupt (or switches-off in the case of the Gō potential) native interactions involving residues located within the knotted core, which otherwise do not play any role in the energetic stabilization (i.e. nucleation) of the folding transition state. An illustrative example is that of a native interaction between a residue located within the threading terminus (TT), with another one located on the knotting loop (KL). SMs are expected to increase the folding rate because they should decrease backtracking by hampering the premature formation of these critical interactions in conformations that still require a substantial amount of structural consolidation. In line with this hypothesis, a significant increase in the folding rate of knotted lattice proteins was observed upon introducing SMs. The observed enhancement was largest (up to 50%) for the $5_2$ knot in comparison with the $3_1$ knot (~36%), indicating that backtracking on lattice polymers is more disruptive in knots of higher complexity (98). This enhancement is particularly striking if one considers the fact that only one TT-KL interaction (i.e. 2% of the total native energy) is being switched-off. The simplicity of lattice models may challenge the relevance of this behavior in real proteins. Nevertheless, it would be interesting to test this prediction in experiments *in vitro.* The importance of the contact map for efficient knotting was addressed in (108) in the context of off-lattice simulations of protein MJ0366.

Topological frustration is expected to lead to folding pathways populated by intermediate states. At the microscopic level, stable non-native interactions, can further contribute to stabilize misfolded intermediates. Therefore, knotted proteins are expected to exhibit complex folding landscapes, populated by intermediate states, which translate into multi-phase kinetics (83,84,101,102,109) and non-linear chevron-behavior (83,95). Recently, single-molecule experiments revealed a rich folding landscape for protein UCH-L1, featuring an ensemble of structurally heterogeneous intermediate states, both on- and off-pathway (106). Some of the intermediate states are mechanically stable and long-lived (i.e. stable for several seconds), and presumably coincide with those previously detected through bulk unfolding studies (109). An ensemble of complex folding pathways with off-pathway misfolded intermediates was also reported for a designed tandem repeat of the hypothetical protein HP0242 (PDB ID: 2bo3 and 4u12), a small (92 residue long) engineered protein (84), which features a knot $3_1$ in its native structure(110).

The folding behavior of knotted proteins is rather different from that exhibited by small-single domain proteins (~100 residues), which is typically well described by a two-state model, and single exponential kinetics (111). Structure-based models, and particularly pure Gō potentials, correctly capture regions of the folding landscape dominated by native interactions. They should be able to predict topologically trapped intermediates, and intermediates states resulting from side-chain packing effects when combined with full atomistic protein representations (49,112,113). Therefore, they are adequate to recapitulate the folding process of model systems with smooth energy landscapes (such as small, single-domain proteins) but will fail to correctly capture regions of the folding landscape dominated by non-native interactions. Given the importance of non-native interactions in the folding of knotted proteins it is important to develop more realistic (though computationally tractable) potentials that will allow to clarify the role of non-native interactions in the energetics, kinetics and thermodynamics of the knotting process.

*Towards a mechanistic understanding of knotting in vivo*

*In vivo*, efficient folding of many newly synthesized proteins is achieved through the utilization of a special class of molecular machines called chaperones, which prevent protein misfolding and aggregation in the crowded environment of the cell (114). The so-called chaperonins comprise an important and universal class of chaperones, which have the unique ability to fold some proteins that cannot be folded by simpler chaperone systems (e.g. the trigger factor Hsp70)(115). Chaperonins are large cylindrical complexes that contain a central cavity that binds to unfolded and misfolded proteins and allows them to fold in isolation **(Fig. 4A-B)**. The most well studied chaperonin is the bacterial GroEL-GroES complex from *E. coli*. (116). During the GroEl-GroES operating cycle, which is driven by ATP, there is a large conformational change that nearly doubles the cavity's volume and is accompanied by a change of the physical properties of its inner walls, which become predominantly hydrophilic and net-negatively charged. The exact mechanism according to which the chaperonin assists folding has not yet been solved, but two main models have been proposed. The passive (or Anfinsen's) cage model predicts that the role of chaperonin is simply that of increasing folding yield by avoiding aggregation. One the other hand, the active cage model envisages an active role for the chaperonin, according to which the latter is able to accelerate folding process by modulating the free energy landscape (116).

A proteome-wide analysis that studied the contribution of the GroEL-GroES complex to the *E. coli*. proteome revealed that the folding process of ~85 of its proteins is stringently chaperon-dependent (117). Interestingly, all these proteins embed a knot in their native structures (118) suggesting that a chaperon-assisted folding is necessary to efficiently fold knotted proteins *in vivo*.

Mallam and Jackson were the first to study the effect of GroEL-GroES on the folding process of *E. coli* proteins YibK and YibA (23). In the presence of GroEL-GroES, folding rate becomes at least 20-fold higher, supporting the view that this mechanism should indeed operate *in vivo*. It was also found that GroEL-GroES has no effect on the folding rate of denatured (and knotted) conformations. Since none of the proteins populates misfolded species, these results indicate that the chaperonin is likely to play a specific and active role in assisting the knotting step. One possibility put forward by the authors is that it drives the formation of knotted chains similar to those that persist in the denatured ensemble upon chemical denaturation.

Inspired by these results Soler *et al.* were the first to use a Gō model to explore via Monte Carlo simulations the folding and knotting mechanism (of a simple cubic lattice with a trefoil knot and a C-alpha off-lattice representation of protein MJ0366) inside the chaperonin cage (119). The chaperonin was represented by a rigid cavity, i.e., the interactions of the amino acids with the confining walls were strictly limited to excluded volume interactions **(Fig. 4C)**. It was found that steric confinement leads to a stabilization of the transition state relative to the unfolded state at the transition temperature, $T_f$, resulting into a faster folding kinetics, in line with previous studies that investigated the role of steric confinement in the folding transition of proteins without knots (120,121). However, it was also reported that steric confinement plays a specific role in the folding of knotted proteins by substantially increasing the knotting probability for high degrees of confinement, even above the melting temperature (i.e. in conditions where the native state is destabilized). In particular, the authors reported that native knotting occurs with moderate probability (>0.5) even before the transition state is crossed, i.e., in conformations with a small degree of structural consolidation. Interestingly, a reduction of backtracking upon steric confinement was also observed, most likely because confinement reduces the frequency of local interactions. Since local interactions contribute to rigidify the protein backbone (i.e. to decrease local flexibility), the authors of (119) performed a systematic study of the effect of local flexibility in the folding process of MJ0366. The conclusion was that when local chain flexibility is increased, MJ0366 samples conformations that are preferably knotted (even slightly above the transition temperature $T_f$), recapitulating the effects of confinement. These results are in line with theoretical analysis, framed on polymer physics considerations, which predicts that local order is an inhibitor of knotting in proteins (9). The impact of chain flexibility on the structure of the transition state ensemble of MJ0366 was also investigated in (119). An enhanced structural flexibility leads to a structural re-arrangement of the transition state, which becomes devoid of helical content (including the C-terminal helix that is threaded through the knotting loop to tangle the protein). Interestingly, the formation of the beta-hairpin, which is the main structural trait of the intermediate slipknotted state identified in previous, full atomistic simulations (113), starts earlier than for stiffer chains, which facilitates the formation of the knot

A follow up study by Niewieczerzal and Sulkowska (118), framed on Molecular Dynamics simulations of a simple C-alpha structured-based Gō model parameterized with the SMOG package (122), extended the study of folding under confinement to other small proteins with a trefoil knot: VirC2 (PDB ID: 2rh3), with the same fold as MJ0366, and DndE (PDB ID: 4lrv), with an unclassified fold, whose knots are only slightly deeper (by 2 and 5 amino acids, respectively) than that of MJ0366. Essentially the same conclusion was recapitulated for the three model systems, namely, that steric confinement upon encapsulation within a cylinder **(Fig. 4D)** facilitates knotting at an early stage of folding, while simultaneously reducing backtracking and smoothing the free energy landscape. However, a new and rather interesting effect of steric confinement was observed when the same simulation approach was used to thoroughly investigate the folding process of protein UCH-L1. As previously outlined, this protein exhibits two folding pathways under bulk conditions. A dominant one in which the N-terminus gets structured first and the protein populates an intermediate conformation with a trefoil knot prior to folding into the native one, and a secondary, less populated pathway, in which the C-terminus gets structured first and the formation of the $5_2$ knot occurs directly from unknotted conformations. What the authors observed was that confinement does not perturb the frequency of the most dominant bulk pathway but leads to a substantial enhancement (e.g. from 2 to 18.5% at high temperature) of the less dominant pathway (99). Furthermore, they noted a significant number of knots (native and non-native) of different types ($3_1$, $4_1$ and $5_2$) in the denatured ensemble of UCH-L1. Surprisingly, the knots are shallow and short lived in contrast with the (less frequent) ones found under bulk conditions.

The results outlined above indicate that steric confinement alone has a direct effect on knotting. By impeding extended conformations, confinement contributes to decrease the entropic cost to knot the chain, facilitating the formation of native knots, even in unfolded conformations. Available data suggests that steric confinement is not able to change the folding pathways observed in the bulk, despite being able to change their frequency. A similar observation holds for the knotting mechanism, which is conserved under steric confinement. In order to broaden and deepen our views on the role of GroEL-GroES on the folding of knotted proteins, more realistic models, encapsulating the physical changes occurring along the chaperonin cycle (123,124) should be applied to knotted proteins. These models may reveal an active role of GroEL-GroES at the level of the folding landscape and knotting mechanism (e.g. by stabilizing key interactions which promote

knotting(24)), which add up to the effects of steric confinement to further increase the folding rate and efficiency.

*Co-translational folding of knotted proteins*

Cieplak and Chwastyk were the first to explore the knotting mechanism of protein YibK during protein synthesis (i.e. co-translationally) via Langevin Molecular Dynamics simulations of a simple C-alpha representation, with protein energetics modeled by a Gō potential (97). Protein synthesis occurs vectorially from the N-terminus to the C-terminus, which remains tethered to the peptidyltransferase center on the larger ribosomal unit (**Fig. 5A**). This implies that if knotting of YibK is to occur co-translationally, it cannot be via a direct threading movement of the C-terminus. As in previous lattice simulations (125), Cieplak and Chwastyk used a minimal representation that reduces the ribosome to a steric plane (**Fig. 5B**). They found that when folding occurs co-translationally, successful knotting (representing 3% of the attempted trajectories) occurs exclusively via slipknotting under optimal temperature conditions. In particular, the degree of confinement provided by the wall facilitates the formation of the C-terminal hairpin on the right side of the knotting loop, while simultaneously assisting threading. These results are interesting because they show that not only it is possible to knot YibK without the assistance of non-native interactions, as they also open-up the possibility that in Lim and Jackson's *in vivo* experiments with YibK fused to ThiS (24) the newly translated proteins were already knotted when folding started.

A subsequent account by Dabrowski-Tumanski *et al*. (126) explored the co-translational knotting mechanism of protein Tp0624 (PDB ID: 5jir) with Molecular Dynamics simulations of a C-alpha model. Tp0624 is the deepest knotted protein found to date, having very long knot tails (with chain length exceeding 120 residues). They used a more sophisticated representation of the ribosome, where the exit tunnel is explicitly represented by a purely steric cylinder; the cylinder is capped at one end by a steric plane mimicking the ribosome's surface. However, in this case, they considered the existence of homogeneous attractive intermolecular interactions between selected beads of the N-terminal domain (e.g. arginines, which are positively charged, are likely to interact with the ribosome) and the confining plane. They found that in up to 60% of the successful folding trajectories, knotting is based on what the authors termed *the ribosome mechanism*. According to the latter, a part of the chain protruding from the tunnel gets wrapped around the exit tunnel and sticks onto the surface. Such a loop is then threaded by the nascent chain that is being pushed out of the tunnel with a constant force. While there are not yet *in vitro* results supporting a role for the ribosome on the knotting mechanism, the results predicted by simulations can, at least partially explain why it is so challenging to knot a protein in molecular simulations.

*Functional advantages of knots in proteins*

Our current understanding of the folding process of knotted proteins shows that these topologically complex molecules are slow folders, which often exhibit relatively complex folding landscapes. In the cell, they are most likely assisted by chaperonins to fold fast and efficiently. This scenario could explain how knotted proteins have withstood evolutionary pressure despite their folding liability (23). However, another possibility is that, to compensate for their slow folding, a knotted native state conveys a functional advantage to the carrier protein. The quest for functional advantages of knots in proteins started with the realization of their existence in the PDB (7), and is also motivated by the observation that knotted motifs are conserved across different families (despite very low sequence similarity)(127).

There is a plethora of studies in the literature, both experimental and computational, which analyzed specific model systems and proposed several biological functions. It has been proposed that knotted proteins have enhanced thermal (20,26), mechanical (26,27,128) and structural (127) stabilities. Another possibility is that knots help shape and stabilize the active site of enzyme (129-131), and in some cases it has even been suggested that knots can control enzymatic (130) and signaling activity (132). A recent study put forward the view that knots can impair protein degradation by ATP-dependent proteases leading to partially degraded products with potentially new functions (133).

By comparing the unfolding rate of $3_1$ and $5_2$ knotted lattice systems with that of their unknotted counterparts Soler *et al.* found a dramatic enhance in kinetic stability for the knotted systems (more than one order of magnitude). Furthermore, it was also observed that kinetic stability of lattice proteins increases considerably with the complexity of knot type (98). An enhanced kinetic stability for knotted proteins was also reported in off-lattice Molecular Dynamics simulations (26,134). A reduction of the intrinsic unfolding rate due to the presence of a knot in the native structure was previously reported by Yeates and co-workers through *in vitro* experiments that also compared an engineered knotted model system with its unknotted counterpart (84). Recently, *in vitro* results showed that a truncated (i.e. lacking eleven N terminal residues) variant of $5_2$ knotted protein UCHL1, unfolds two-orders of magnitude faster than the full UCHL1(27). Since the truncated variant is unknotted, this finding also supports the view that knots enhance kinetic stability.

Kinetic stability is a property of the native state that is essential to maintain the biological function of the protein during a physiologically relevant timescale (135), and therefore may represent a functional advantage. An enhanced kinetic stability may be particularly advantageous for proteins forming transmembrane channels since they are subjected to mechanical stress. Since knotted patterns appear to be preferentially conserved among transmembrane channels (127) we propose, as Faísca did in (22), that a *systematic* functional role of knots in proteins is precisely that of enhancing their kinetic stability.

*Conclusions and Future prospects*

Interest in the field of knotted proteins has been increasing over the last decade, with many important contributions (both theoretical and experimental) from several research groups worldwide. This expansion contributed to significantly advance our understanding of their intricate knotting mechanism, with experiments corroborating previous theoretical predictions for specific knotted proteins. While focus has mostly been placed on proteins embedding knotted trefoils, some recent studies featured model systems with knot types of higher complexity. This is, in our view, a line of research that is worth pursuing as an attempt to establish the different types of knotting mechanisms, how much they depend on knot complexity, how are they conserved across different knotted proteins, and how they affect the overall folding performance.

Microscopically, it is extremely important to use molecular simulations to understand and determine the exact role played by non-native interactions in the kinetics, energetics and thermodynamics of knotting. We also envisage an important role for molecular simulations in what regards the determination of the knotting mechanism inside the “chamber of secrets” (115), for which direct experimental observations are not yet possible. Current contributions focused on the steric effects played by the chaperonin’s chamber, but it is crucial to use more realistic models to that take into account the physical changes that take place during the working cycle, and which may play an active role in the physics of knotting. Is the enhanced knotting frequency that occurs under steric confinement conserved when other intermolecular interactions (e.g. hydrophobic and electrostatic) are included? Can the latter modulate the free energy landscape so that productive intermediate states, with enhanced knotting frequency, become populated?

Establishing the role of chaperonin-assisted knotting will not only enlarge our vistas on chaperon Biology, but it will also contribute to clarify our understanding of the evolution of knotted proteins and, consequently, of their functional role.

## Acknowledgments

Work supported by UID/MULTI/04046/2013 centre grant from FCT, Portugal (to BioISI). PFNF would like to acknowledge the contribution of the COST Action CA17139.

**Figures**

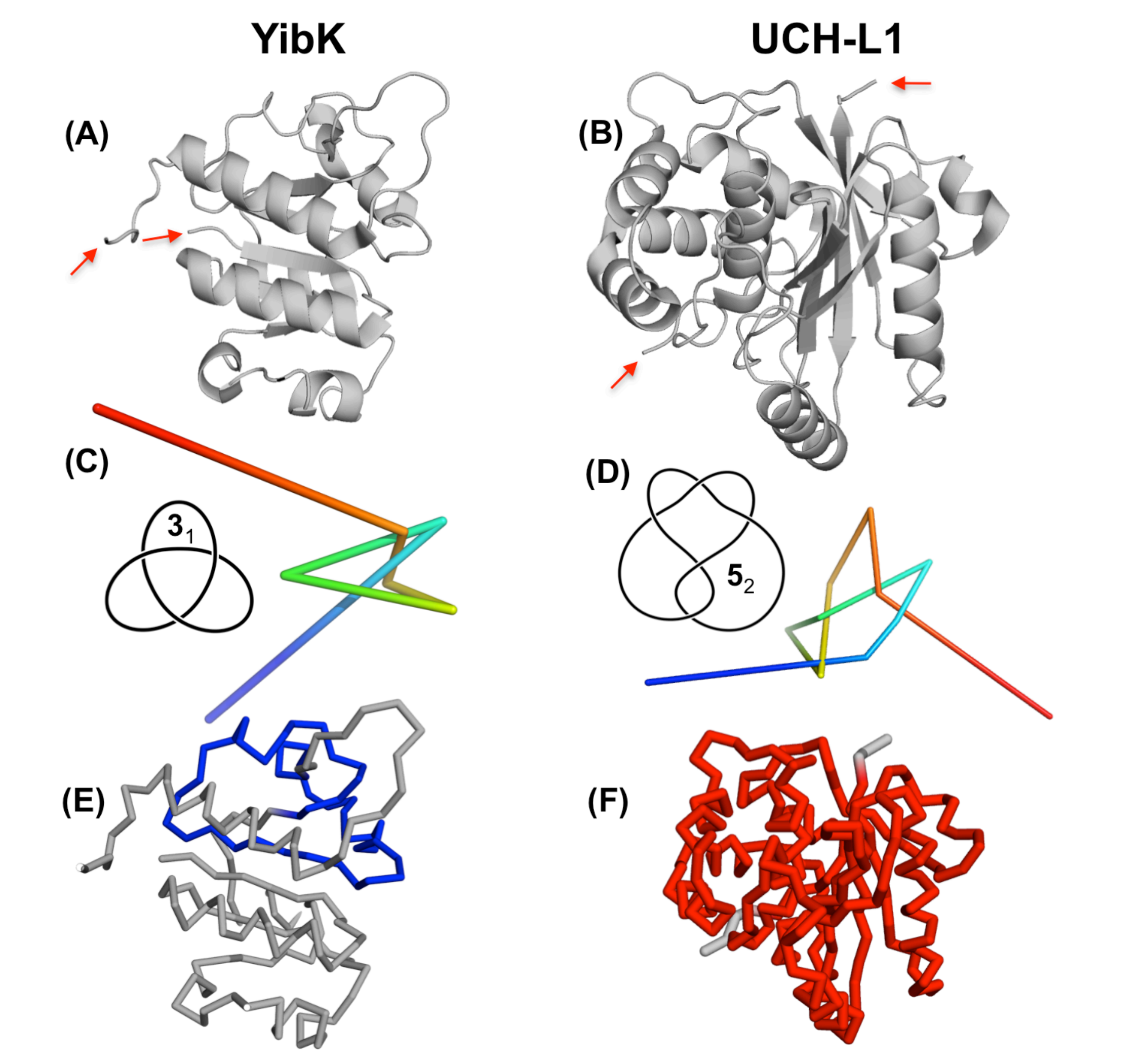

**Figure 1**. Three-dimensional cartoon representation of the native structure of protein YibK (A) and UCH-L1 (B). Note that the termini (denoted by arrows) are located very close together in the case of YibK, and connecting them such that the backbone makes a closed curve is space is also straightforward for UCH-L1. Reduced representation of YibK (C) and UCH-L1 (D) resulting from the application of the KMT algorithm, which highlights the existence of a knot $3_1$ in YibK and knot $5_2$ in UCH-L1. Ribbon representation with the knotted core (i.e. the minimal part of the chain that contains the knot) highlighted in blue (E) and red (F). In the case of YibK both knot tails are long and the knot classifies as deep, while for UCH-L1 they are short and the knot classifies as shallow.

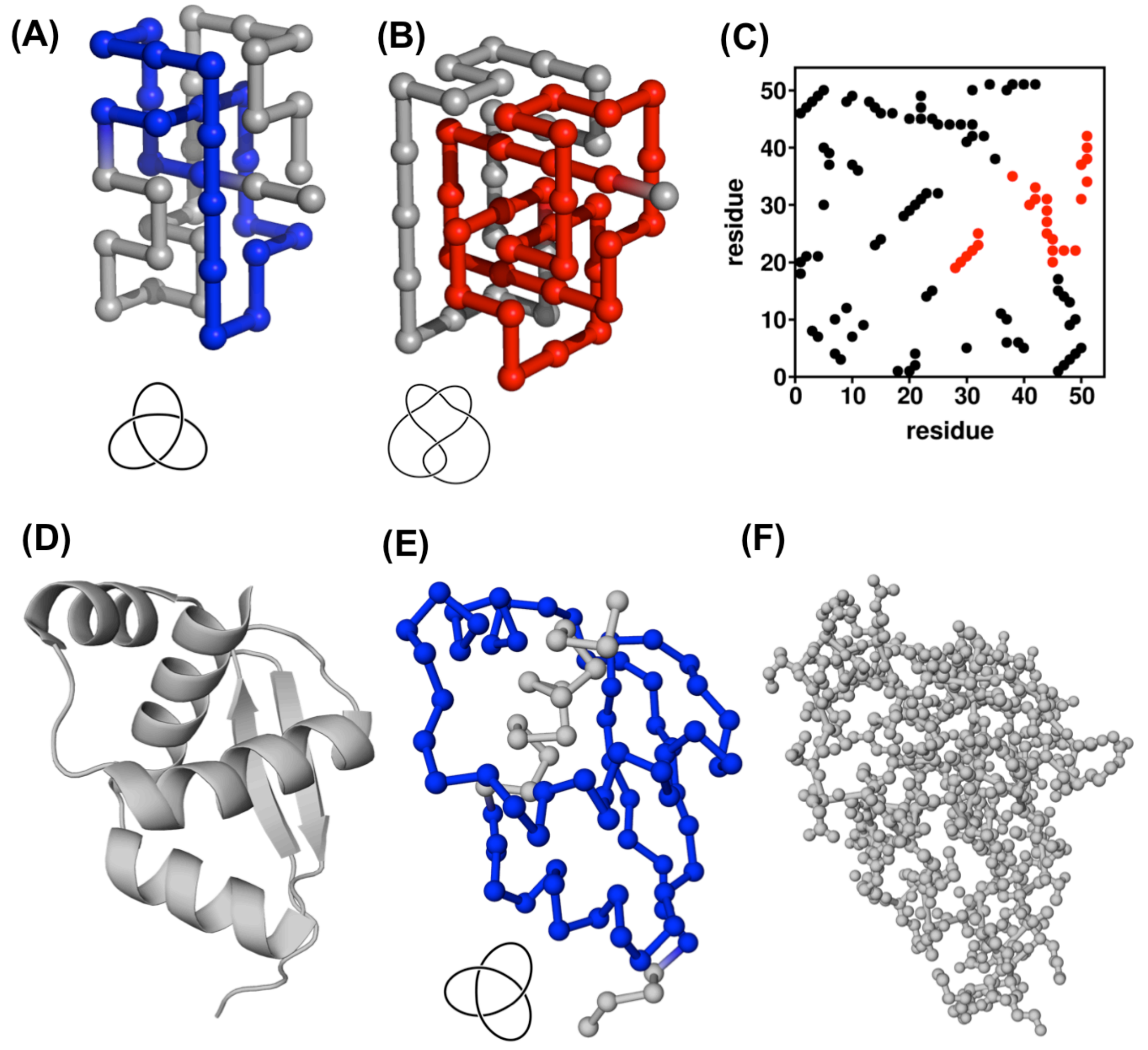

**Figure 2**. Simple lattice representations of a shallow knotted trefoil (A), and of a shallow knotted twisted-three knot (B), both designed by hand, used in Monte Carlo simulations. The lattice representation reduces the amino acids to beads of uniform size, and restricts them to occupy the vertices of a regular lattice. The knotted core is highlighted in blue (knot $3_1$) and red (knot $5_2$). Contact map (C) representing the set of 52 intra-molecular interactions that exist in the native conformation. On-lattice it is straightforward to define a native interaction: Is one that establishes between beads that are separated by one lattice spacing, but are not linked along the protein's backbone. In a structure-based model, like the Gō model, only native interactions contribute to stabilize the protein as it folds towards the native conformation. In the contact map the red points represent interactions that establish between beads pertaining to the knotted core. Three-dimensional cartoon representation of the native structure of MJ0366, the smallest knotted protein found in the PDB, which embeds a trefoil knot (D). C-alpha continuum representation of MJ0366 (E) where each amino acid is reduced to a bead of uniform size centered at the respective C-alpha atom. Beads are connected along the chain by sticks representing the peptide bond.

The knotted core is highlighted in blue in the C-alpha representation. Since one of the knot's tails only spans four residues, the knot classifies as shallow. The most complex representation of the native structure is the full atomistic representation, where all the atoms of the protein (except hydrogen) are explicitly represented (F).

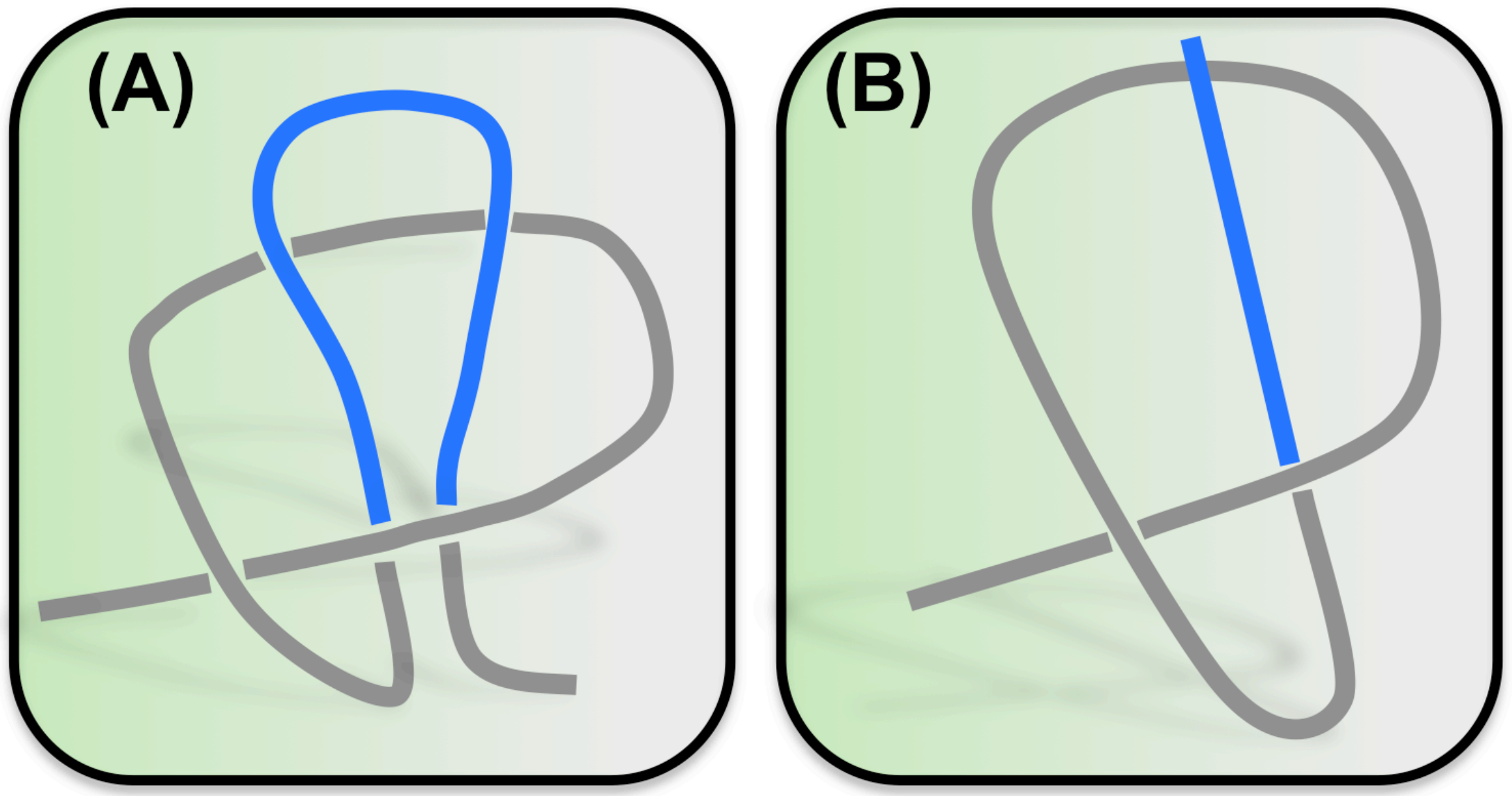

**Figure 3.** Structure of the knotting step as predicted by simulations. In splipknotting (A) one of the chain termini arranges itself in a hairpin like conformation that threads the knotting loop formed by the remainder of the chain. An alternative mechanism involves direct threading of the termini through the knotting loop (B).

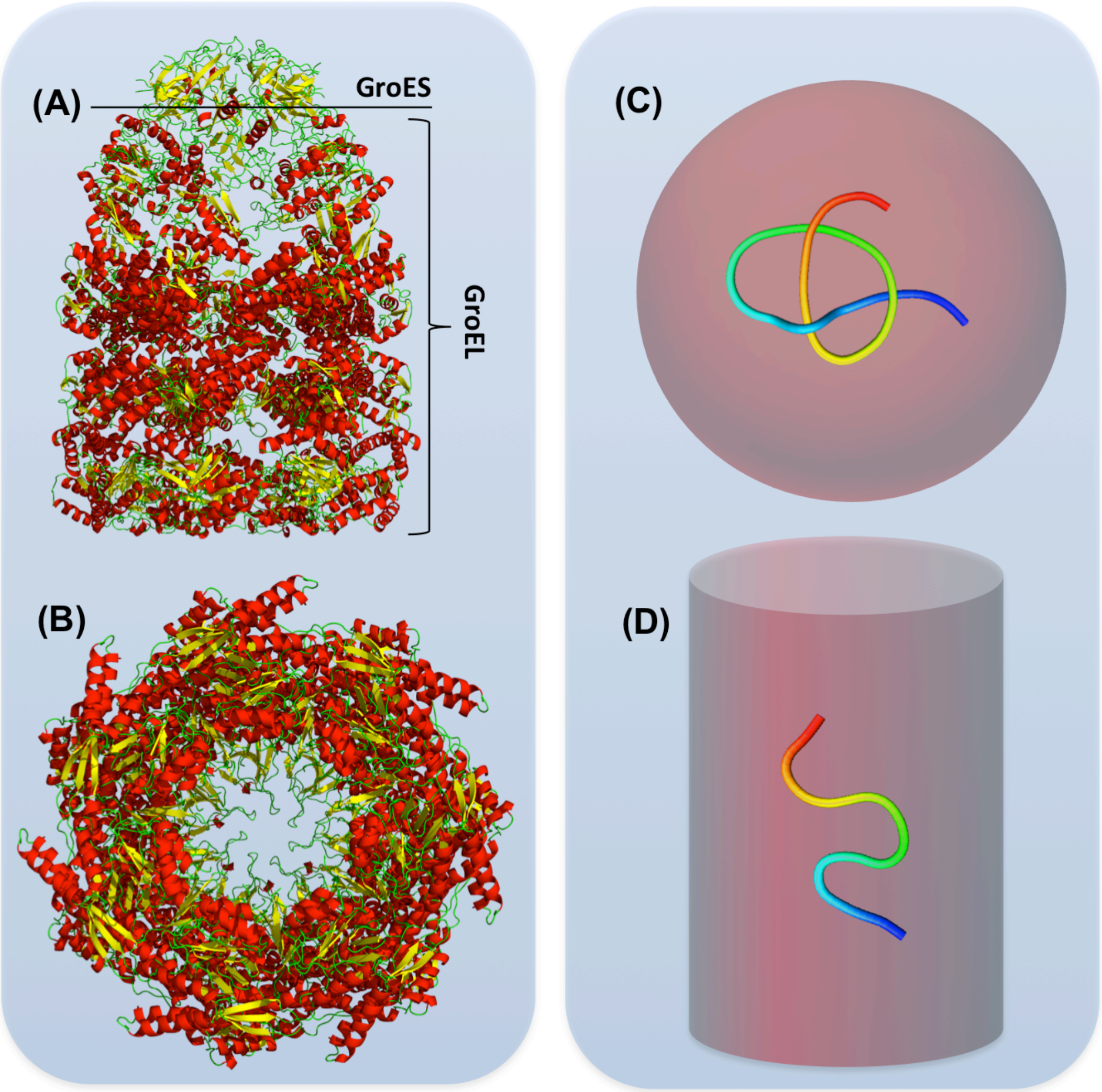

**Figure 4.** Three-dimensional structure of the GroEL-GroES complex (PDB ID: 1aon) (A), and end view of the GroEL-GroES, highlighting the central cavity where folding takes place (B). Simple models of steric confinement used to mimic excluded volume interactions between the protein and the chaperonin chamber. A spherical shape (C) was adopted in (119), while a cylindrical (D) one was adopted in (118).

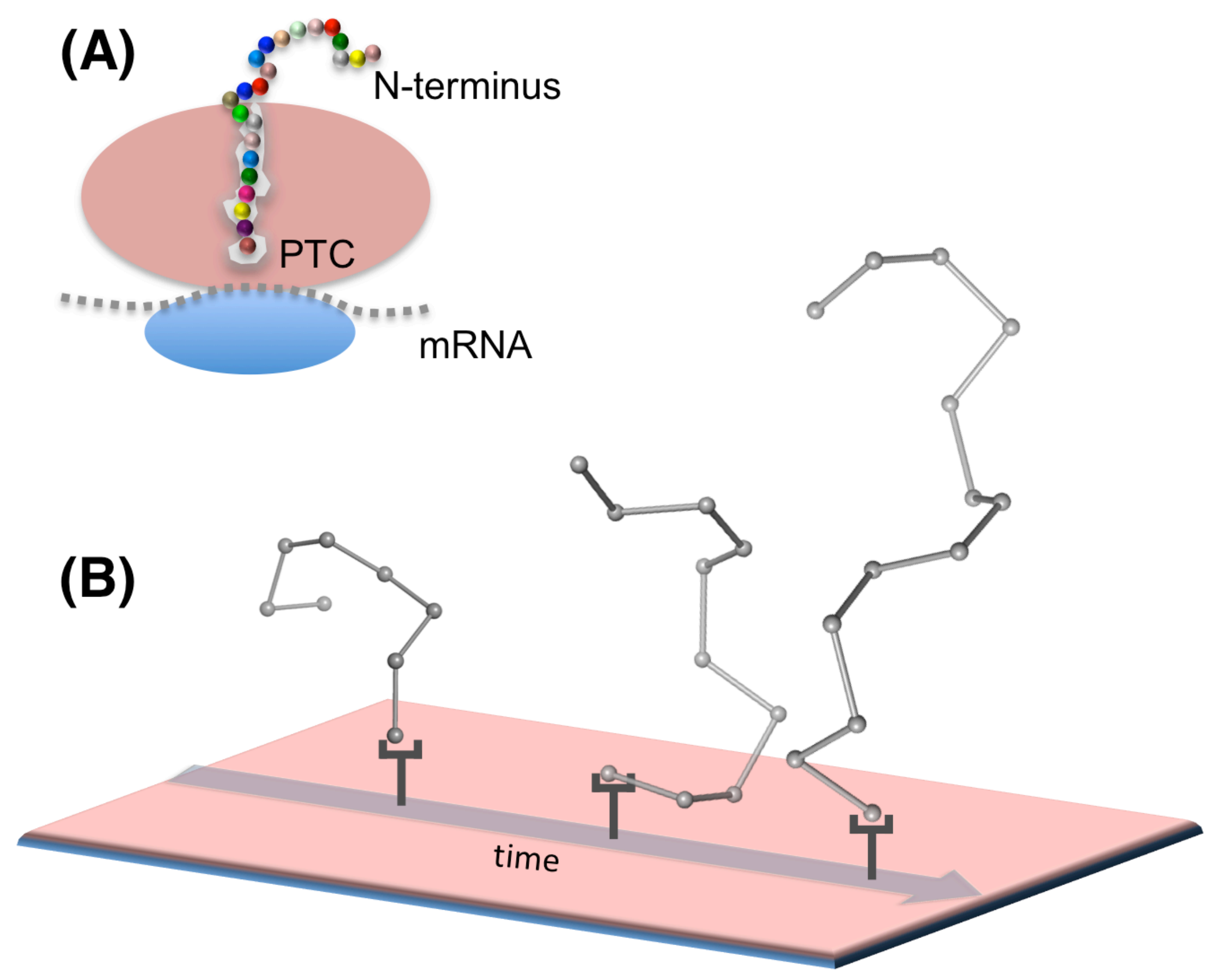

**Figure 5.** Cartoon representation of the ribosome highlighting the exit tunnel (located on on the larger ribosomal unit), and the vectorial nature of protein synthesis. The latter which proceeds from the N-terminus to the C-terminus that is tethered at the PTC on the larger ribosomal unit (A). Simulation set-up in which the larger ribosomal unit is represented by a steric plane, and the protein (represented by a C-alpha model), is tethered to the plane by the C-terminus (B). Adapted from (125)